# Hybrid Fiber-Based Radio Frequency Distribution and Vibration Detection System Tailored for Large Radio Arrays

HONGFEI DAI,[1,2] WENLIN LI,[1,2] ZHONGWANG PANG,[1,2] CHUNYI LI,[1,2] DONGQI SONG,[1,2] TONG WU,[1,2] AND BO WANG[1,2,*]

[1]*State Key Laboratory of Precision Space-time Information Sensing Technology, Department of Precision Instrument, Tsinghua University, Beijing 100084, China*
[2]*Key Laboratory of Photonic Control Technology, Ministry of Education, Tsinghua University, Beijing 100084, China*
*bo.wang@tsinghua.edu.cn

**Abstract:** Radio telescope arrays, such as Square Kilometre Array (SKA) and next-generation Very Large Array (ngVLA), require highly precise synchronization of time-frequency references to ensure high-quality observational data. Fiber-based frequency distribution systems are highly effective. However, their proper functioning can be threatened by risk events. In this paper, we propose a hybrid fiber-based frequency-distribution and vibration detection system tailored for large radio arrays. The system ensures the performance of distributed frequency signals while allowing for the monitoring of potential threats to the optical fiber network. We design and implement a single-to-multiple hybrid system, conducting tests via a 55-km fiber link. Experimental results demonstrate its effectiveness, achieving the relative frequency stability of $3.0\times10^{-14}/1$ s and $2.7\times10^{-17}/10^5$ s, along with vibration detection and localization capabilities.

## 1. Introduction

Radio telescope arrays can observe radio signals from the universe through the combined operation of multiple telescopes, supporting humanity in addressing significant astronomical questions [1-5]. Typical examples of large radio telescope arrays include Square Kilometre Array (SKA) [6] and next-generation Very Large Array (ngVLA) [7]. These arrays consist of thousands of receiving antennas. To correctly synthesize the signals from receiving antennas, frequency references of each receiver in a large array must be precisely synchronized. Otherwise, phase errors will occur between the observed signals, reducing the quality of the observational data [8].

Among various frequency distribution solutions, fiber-based frequency distribution stands out due to their reliability and cost-effectiveness [9-17]. Several radio-frequency over fiber (RFoF) solutions have been used in radio telescope arrays [18-20]. These solutions efficiently compensate for the phase noise introduced during the transmission of frequency signals, which is caused by factors such as temperature fluctuations and vibrations. As a result, they can disseminate signals from atomic clocks, such as hydrogen maser, at the central station to remote receiving stations with high quality. As a critical foundation for the array, maintaining the fiber-based frequency distribution network is essential.

Large-scale fiber networks may face various threats, such as weather, animals, human activities, and natural disasters. Although fiber-based frequency distribution can reduce the negative impacts of most reversible events, it has its limitations. In particular, some risks can cause irreversible and destructive damage. Several studies have shown that tampering with fiber-based time-frequency distribution is possible, and it is difficult to detect [21-23]. Moreover, fiber networks are also responsible for data transmission and inter-station communication. It also highlights the importance of protecting these fiber networks. However, fiber networks specifically built for radio arrays are large in scale, making traditional risk monitoring methods, such as dedicated sensors and video monitoring, unaffordable. Therefore, it is crucial to choose suitable risk monitoring methods to detect and locate these risks.

Distributed optical fiber sensing (DOFS) is suitable for this scenario. The risk events mentioned earlier, which can cause destructive impacts on fiber networks, are often preceded by abnormal vibrations. They can be detected by DOFS systems. Furthermore, vibration detection enables more scientific functions, such as seismic detection, ocean monitoring, and traffic flow analysis [24-27], thereby maximizing the overall scientific value of the radio array. DOFS methods for vibration detection can be divided into two types. One is based on backscatter-based structures, such as distributed acoustic sensing (DAS) [28] and Brillouin optical time domain reflectometry (B-OTDR) [29]. The other is based on forward interferometric structures [24, 30]. In fact, fiber-based frequency distribution systems also base on building radio-frequency interferometers to detect and compensate for phase noise introduced by the fiber link. It means that DOFS systems based on forward interferometric structures can theoretically integrate well with fiber-based frequency distribution systems, providing both functions in a single system.

Previously, Sebastian Noe proposed a method combining optical frequency transmission and seismic detection [31], but it lacks the ability to accurately locate vibrations. Adonis Bogris introduced a sensitive sensing method based on microwave frequency fiber interferometry [32], but this method does not support frequency distribution, and it also lacks the ability to accurately locate vibrations. These methods represent initial explorations of multifunctional sharing of fiber networks. However, they have practical limitations that make it challenging to effectively apply them to solve frequency distribution issues and address link risks.

In this paper, we propose a hybrid fiber-based radio frequency distribution and vibration detection system, which is suitable for large radio arrays. This system meets the frequency distribution requirements between a central station and multiple remote stations. While maintaining the performance of frequency distribution, the optical signals transmitted through the fiber link are reused to simultaneously enable vibration detection and localization. In the experimental demonstration, we achieved frequency distribution with the relative frequency stability of $3.0\times10^{-14}/1$ s and $2.7\times10^{-17}/10^5$ s via a 55-km fiber. Meanwhile, the hybrid system can accurately detect and locate vibrations occurring on the link. It offers a potential solution for enhancing the security of large radio arrays and can be used in other scenarios that rely on fiber networks for frequency distribution.

## 2. Method

Long-distance transmission will introduce unwanted phase noise to the frequency signal, which can worsen its performance. Therefore, the primary challenge in fiber-based frequency transmission lies in compensating for the fiber-induced phase noise. Moreover, a central station has to concurrently support the frequency distribution requirements of multiple remote stations for radio arrays. To address these requirements, based on [18, 33], we propose the hybrid system, as shown in Fig. 1(a). We denote the RF (radio frequency) signal at the remote processing facility (RPF) as $V_R(f_R, t) = cos(2\pi f_R t + \varphi_R)$. For convenience, the amplitude of these RF signals is disregarded. The RPF modulates $V_R(f_R, t)$ onto an optical carrier, whose wavelength is $\lambda_1$, and then transmits the modulated signal to the central processing facility (CPF). After transmission through the optical fiber link, the optical signal received by the CPF will have one-way phase noise $\varphi_n$. We denote the received RF signal as $V_R'(f_R, t) = cos(2\pi f_R t + \varphi_R + \varphi_n)$. Subsequently, $V_R'(f_R, t)$ is modulated onto another optical carrier, whose wavelength is $\lambda_2$, and then transmitted back to the RPF. The use of a different optical carrier wavelength is to mitigate the effects of Rayleigh scattering on the transmission of $V_R(f_R, t)$.

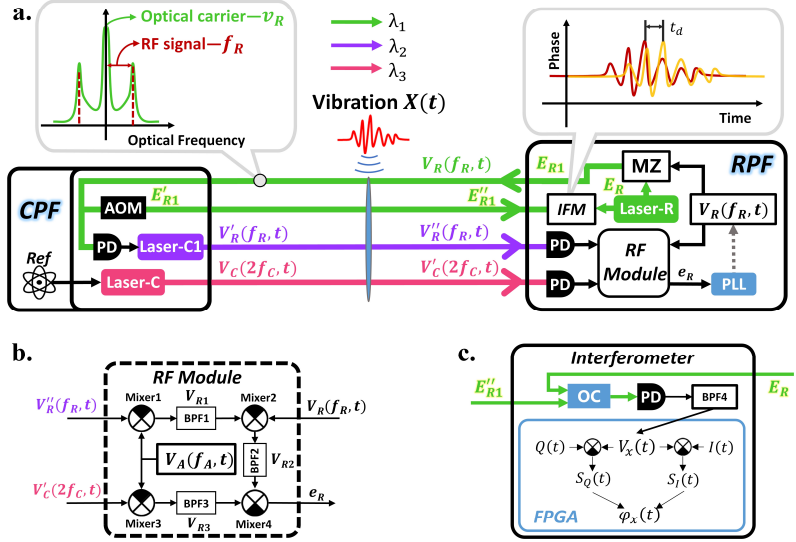

Fig. 1. (a) Schematic diagram of the hybrid fiber-based frequency-distribution and vibration detection system. (b) Details of the RF module. (c) Details of the interferometer and phase calculation flowchart. AOM: acousto-optic modulator, PD: photodetector, RF: radio frequency, IFM: interferometer, MZ: Mach-Zehnder modulator, PLL: phase-locked loop, BPF: bandpass filter, CPF: central processing facility, RPF: remote processing facility.

After completing this loop, the frequency signal received and demodulated by the RPF is

$$V_R''(f_R,t) = cos(2\pi f_R t + \varphi_R + 2\varphi_n). \tag{1}$$

Meanwhile, on the CPF side, the reference frequency signal is upconverted to $V_C(2f_C,t) = cos(2\pi \cdot 2f_C t + 2\varphi_C)$. The frequency $f_C$ and $f_R$ are initially set to be very close. At the CPF, $V_C(2f_C,t)$ is modulated onto another optical carrier, whose wavelength is $\lambda_3$, and transmitted to RPF. Since the frequency of $V_C(2f_C,t)$ is about twice that of $V_R(f_R,t)$, this one-way transmission introduces phase noise of $2\varphi_n$. It means that the signal received by the RPF is

$$V_C'(2f_C,t) = cos(2\pi \cdot 2f_C t + 2\varphi_C + 2\varphi_n). \tag{2}$$

Theoretically, by mixing $V_R''(f_R,t)$ and $V_C'(2f_C,t)$ followed by bandpass filtering, the link fluctuation $\varphi_n$ can be eliminated. However, in practice, due to signal leakage from the mixer components, some noise signals originating from $V_R''(f_R,t)$ will leak into subsequent phase-locked processing steps, leading to incomplete elimination of the phase noise. To overcome this issue, we design a leakage-guard RF module shown in Fig. 1(b). This module introduces an auxiliary frequency source, i.e., $V_A(f_A,t) = cos(2\pi f_A t + \varphi_A)$. The choice of frequency $f_A$ should be such that it is distinguishable from the frequency signals generated by nonlinear effects or leakage, specifically those generated by $V_R''(f_R,t)$ and $V_C'(2f_C,t)$.

Subsequently, we perform several mixing steps. First, $V_R''(f_R,t)$ is mixed with $V_A(f_A,t)$ passed through a bandpass filter to obtain the signal $V_{R1}$, i.e.,

$$V_{R1} = cos[2\pi \cdot (f_R + f_A)t + \varphi_A + \varphi_R + 2\varphi_n]. \tag{3}$$

Then, $V_{R1}$ is mixed with $V_R(f_R,t)$ passed through a bandpass filter to produce the signal $V_{R2}$, i.e.,

$$V_{R2} = cos[2\pi \cdot (2f_R + f_A)t + \varphi_A + 2\varphi_R + 2\varphi_n]. \tag{4}$$

Meanwhile, $V_C'(2f_C,t)$ is mixed with $V_S(f_S,t)$ passed through a bandpass filter to obtain the signal $V_{R3}$, i.e.,

$$V_{R3} = cos[2\pi \cdot (2f_C + f_A)t + \varphi_A + 2\varphi_C + 2\varphi_n]. \tag{5}$$

Then, $V_{R2}$ is mixed with $V_{R3}$ to produce the error signal $e_R$, i.e.,

$$e_R = cos[2\pi \cdot 2(f_R - f_C)t + 2(\varphi_R - \varphi_C)]. \tag{6}$$

Based on $e_R$, phase locking is achieved using a servo controller, allowing adjustment of $V_R(f_R, t)$ to satisfy the following relationship:

$$f_R - f_C = 0, \tag{7}$$

$$\varphi_R - \varphi_C = 0. \tag{8}$$

At this point, the system mitigates the impact of signal leakage in the frequency distribution system, and frequency distribution between the CPF and RPF is achieved.

Next, we analyze how vibration detection and localization are achieved within this hybrid system. The optical signal transmitted from the RPF is defined as $E_{R1}$, which has already been modulated, i.e.,

$$E_{R1} = V_R(f_R, t) \cdot cos(2\pi v_R t + \varphi_v), \tag{9}$$

where $v_R$ is the optical frequency of Laser-R, and $\varphi_v$ is the initial phase. The performance of the laser can affect the vibration detection performance. This impact can be minimized by selecting an appropriate laser, which has been thoroughly discussed in previous research [34]. After modulation, the optical signal transmitted from the RPF exhibits sidebands in its frequency spectrum as shown in Fig. 1(a). The sidebands are used to carry the disseminated frequency, while the initial optical carrier enables vibration detection.

If a vibration $X(t)$ occurs at a position $L_x$ away from RPF, it will affect the phase of the optical signal in transmission. Upon reaching the CPF, a part of received optical signal passes through an acousto-optic modulator (AOM), generating a frequency-shifted signal with an offset of $f_{AOM}$, i.e.,

$$E'_{R1} = V'_R(f_R, t) \cdot cos[2\pi(v_R + f_{AOM})t + \varphi_v + aX(t)], \tag{10}$$

where $a$ is the vibration coupling factor. Then, $E'_{R1}$ is sent back to RPF. On the return path, the vibration $X(t)$ will further influence the phase of the optical signal, causing the optical signal arriving at the RPF to become $E''_{R1}$, i.e.,

$$E''_{R1} = V''_R(f_R, t) \cdot cos[2\pi(v_R + f_{AOM})t + \varphi_v + aX(t) + aX(t + t_d)], \tag{11}$$

where $t_d$ represents the time difference between the first and second instances when the vibration $X(t)$ affects the optical phase. At the RPF, $E''_R$ interferes with the initially unmodulated optical signal $E_R = cos(2\pi v_R t + \varphi_v)$, using a photodetector (PD) for detection. The detected signal is $V_x(t)$, i.e.,

$$V_x(t) = A(t) \cdot cos[2\pi f_{AOM}t + aX(t) + aX(t + t_d)]. \tag{12}$$

where $A(t)$ represents the amplitude. We use a data acquisition system to collect $V_x(t)$ and perform subsequent processing in the field-programmable gate array (FPGA). FPGA generates the digital quadrature signals $Q(t)$ and $I(t)$, i.e.,

$$Q(t) = cos(2\pi f_{AOM}t), \tag{13}$$

$$I(t) = sin(2\pi f_{AOM}t). \tag{14}$$

$Q(t)$ and $I(t)$ are mixed with $V_x(t)$ and then low-pass filtered, resulting in $S_Q(t)$ and $S_I(t)$,

$$S_Q(t) = A(t) \cdot cos[aX(t) + aX(t + t_d)], \tag{15}$$

$$S_I(t) = A(t) \cdot sin[aX(t) + aX(t + t_d)]. \tag{16}$$

By applying the arctangent calculation method, we can extract the phase $\varphi_x(t)$,

$$\varphi_x(t) = aX(t) + aX(t + t_d) = arctan[S_I(t)/S_Q(t)]. \tag{17}$$

This phase calculation method can avoid the impact of amplitude fluctuations in the interference signal $V_x(t)$. By analyzing $\varphi_x(t)$, we can determine whether there are any abnormal vibrations along the link. For clarity, the phase information shown in Fig. 1(a) corresponding to $X(t)$ and $X(t + t_d)$ is represented by the yellow and red lines, respectively. In fact, the detected phase signal $\varphi_x(t)$ is the superposition of $X(t)$ and $X(t + t_d)$ cannot be directly separated. Therefore, it is necessary to analyze $\varphi_x$ to extract $t_d$. In practice, $t_d$ is extremely small. For a 100 km optical fiber, it does not exceed 1 ms. This makes it difficult to accurately compute the time delay using cross-correlation algorithms, especially for low-frequency vibrations.

In this paper, we use the mirror-image correlation method for localization to figure out $t_d$ of the vibration event [35]. Next, we briefly introduce the mirror-image correlation method. The input to the correlation method is $\varphi_x(t)$. We start by choosing an initial time delay $\tau$. The input $\varphi_x(t)$ is then delayed by $\tau$ and multiplied by $(-1)^1$, resulting in the first-order mirrored signal $S_1(\tau)$,

$$S_1(\tau) = (-1)^1 \varphi_x(t + \tau) = (-1)^1[aX(t + \tau) + aX(t + t_d + \tau)]. \tag{18}$$

Then, $S_1(t_0)$ is added to $S_0(t)$, resulting in the first-order error signal $\delta_1(t_0)$

$$\delta_1(\tau) = aX(t) - [aX(t + \tau) - aX(t + t_d)] - aX(t + t_d + \tau). \tag{19}$$

Here, we can observe that if $\tau$ does not match $t_d$, $\delta_1(\tau)$ will contain a corresponding error term, i.e., $[aX(t + \tau) - aX(t + t_d)]$.

Next, we delay $\varphi_x(t)$ by $2\tau$ and multiplied by $(-1)^2$, resulting in the second-order mirror signal $S_2(\tau)$,

$$S_2(\tau) = (-1)^2 \varphi_x(t + 2\tau) = (-1)^2[aX(t + 2\tau) + aX(t + t_d + 2\tau)]. \tag{20}$$

Then, $S_2(\tau)$ is added to $\delta_1(\tau)$, resulting in the second-order error signal $\delta_2(\tau)$,

$$\delta_2(\tau) = aX(t) - [aX(t + \tau) - aX(t + t_d)] + [aX(t + 2\tau) - aX(t + t_d + \tau)] \tag{21}$$
$$+ aX(t + t_d + 2\tau).$$

The process from Eq. (18) to (21) is repeated. As a result, we obtain the $m$-th order error signal $\delta_m(\tau)$,

$$\delta_m(\tau) = aX(t) + (-1)^m aX(t + t_d + m\tau) \tag{22}$$
$$+ \sum_{i=1}^{m}(-1)^i \{aX(t + i\tau) - aX[t + t_d + (i - 1)\tau]\}.$$

Compared to $\varphi_x(t)$, $\delta_m(\tau)$ expands the time delay between the two vibration signals from $t_d$ to $(t_d + m\tau)$, introducing $m$ error terms. The smaller these error terms are, the closer $\tau$ is to $t_d$. To quantify the proximity between $\tau$ and $t_d$, we need to construct an indicator.

The periodicity of the variation in these error terms of $\delta_m(\tau)$ is $2\tau$. Therefore, the power of the $\frac{1}{2\tau}$ frequency component can be used to represent the magnitude of the error terms, i.e.,

$$P(\frac{1}{2\tau}) = \left|\tilde{\delta}_m(\frac{1}{2\tau})\right|^2. \tag{23}$$

Since the value of $P(\frac{1}{2\tau})$ does not have physical significance and is only used for comparison, we normalize it to obtain the localization indicator function, i.e., power of mark error $M(\tau)$,

$$M(\tau) = 10lg\left[\frac{P\left(\frac{1}{2\tau}\right)}{\int_0^{t_{max}} P\left(\frac{1}{2\tau}\right)}\right]. \tag{24}$$

To accurately get $t_d$, the value of $\tau$ is varied between 0 and the maximum time delay $t_{max}$, where $t_{max}$ represents the time required for the complete round-trip of the optical signal through the fiber link. When the value of $\tau$ is closest to $t_d$, $M(\tau)$ will reach its minimum. This localization method has the characteristic that higher frequency vibration events result in better localization performance. This is because, over the same time interval, higher frequency vibrations generate larger error terms within $\delta_m(\tau)$, making the minimum value of $M(\tau)$ easier to distinguish [35]. Previous research has demonstrated that the frequencies of vibrations induced by human activities and animal activities on optical fibers are primarily distributed between 10 Hz and 40 Hz [28]. It highlights the significance of validating the ability of the system to detect low-frequency vibrations. Details of this will be described in the experiment section.

As for the resolution of localization, it largely depends on the sampling rate $f_s$. It corresponds to the minimum step value $\Delta\tau_{min}$ during the traversal of $\tau$ in the localization method, i.e.,

$$\Delta\tau_{min} = \frac{1}{f_s}. \tag{25}$$

After obtaining the accurate time delay $t_d$, it indicates the location where the vibration occurred, as the following relationship:

$$L_x = L - \frac{ct_d}{2n}, \tag{26}$$

where $c$ is the speed of light in the vacuum, $n$ is the refractive index of the fiber, and $L$ is the total length of the fiber.

## 3. Experiment

### 3.1 Experimental setup

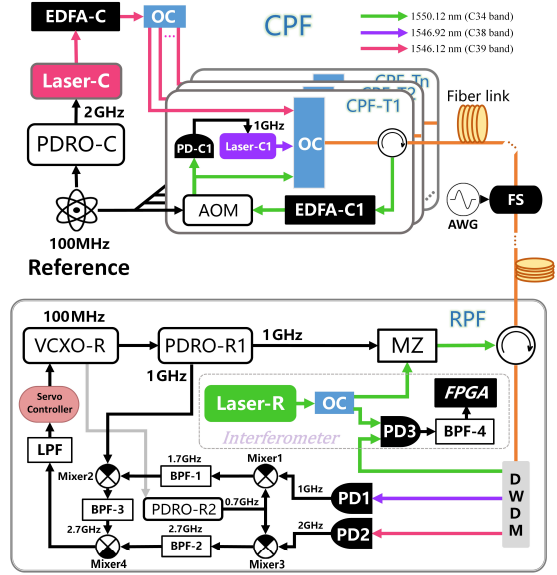

Fig. 2. Details of the hybrid fiber-based frequency-distribution and vibration detection system, including a CPF and an RPF. PDRO: phase-locked dielectric resonant oscillator, EDFA: erbium-

doped fiber amplifier, OC: optical coupler, FS: fiber stretcher, BPF: band-pass filter, LPF: low-pass filter, VCXO: voltage-controlled crystal oscillator, DWDM: dense wavelength division multiplexing, AWG: arbitrary waveform generator. FPGA: field-programmable gate array.

We construct an experimental system as shown in Fig. 2. The part of the CPF presets multiple transmitting cards. Each transmitting card supports one RPF in meeting the frequency distribution requirements. The other part of the CPF only needs to provide each transmitting card with the laser from Laser-C and the frequency reference. This design allows the CPF to be flexibly adjusted based on actual needs. Laser-C and Laser-C1 are distributed feedback (DFB) lasers with a linewidth of around 1 MHz. The wavelength of Laser-C is 1546.12 nm (C39 band), while the wavelength of Laser-C1 is 1546.92 nm (C38 band). The output optical power of Laser-C and Laser-C1 is approximately 3 dBm. The 100-MHz frequency reference is sourced from a Hydrogen Maser. In the actual experiment, we conduct testing and validation using one of the transmitting cards.

At the RPF end, Laser-R is an NKT Koheras BASIK X15, with a linewidth of less than 0.1 kHz. Its wavelength is 1550.12 nm (C34 band) and its output optical power is about 3 dBm. Mixer-1, Mixer-2, and Mixer-3 are used for up-conversion frequency mixing. Mixer-4 is used for down-conversion frequency mixing. BPF-1, BPF-2, BPF-3, and BPF-4 each have a center frequency around 1.7 GHz, 2.7 GHz, 2.7 GHz, and 100 MHz, respectively. RPF is connected to the CPF via a 55-km fiber, i.e., $L$ =55.560 km. This length is measured in advance using an optical time domain reflectometer (OTDR, YOKOGAWA, AQ7275). The type of fibers used in the experiment is G652.D, whose refractive index $n \approx 1.467$. Additionally, a fiber stretcher (FS) is inserted at different locations of fiber link to generate vibrations for subsequent tests. At the RPF end, the sampling frequency $f_s$ for $V_x(t)$ is 5 MHz.

### 3.2 Experimental results

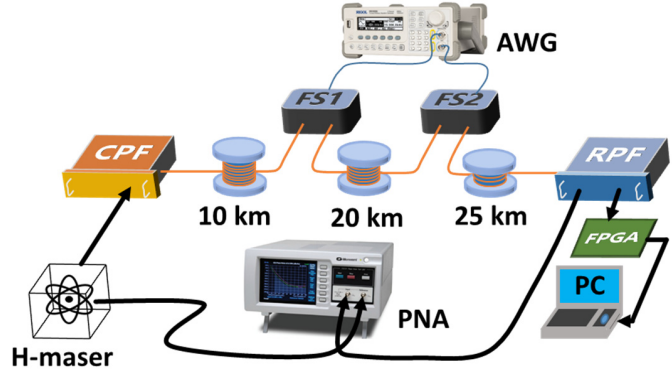

Fig. 3. Schematic diagram of the experimental system and test equipment. PNA: phase noise analyzer, PC: personal computer, AWG: arbitrary waveform generator, H-maser: Hydrogen maser. The orange line represents the fiber link.

The experimental system and test equipment are shown in Fig. 3, and all of them are placed within the same laboratory. For frequency distribution, the equipment used to measure frequency stability and phase drift is the phase noise analyzer (PNA, Microsemi, 5125A). The 100-MHz signal from a hydrogen atomic clock serves as the reference for CPF and PNA. The 100-MHz frequency signal disseminated through the 55-km fiber link is then compared with the reference signal using the PNA. For vibration detection, FPGA is used for data acquisition and phase data extraction, which is then transmitted to a PC for vibration localization algorithms. The arbitrary waveform generator (AWG, RIGOL, DG1022U) generates corresponding sine wave signals and applies them to two fiber stretchers (FS, General Photonics, FST-001-B) to generate the vibrations.

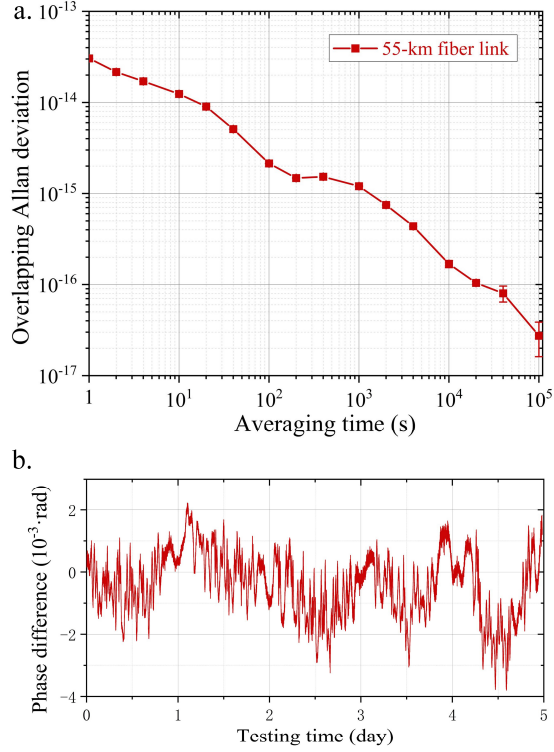

Fig. 4. Measured relative frequency stability and phase drift between the 100-MHz frequency reference at the CPF and the synchronized 100-MHz signal at the RPF end. (a) Relative frequency stability results on the 55-km fiber link. (b) Phase drift over a 5-day testing period.

The relative stability test results are shown in Fig. 4(a). The relative frequency stability reached $3.0 \times 10^{-14}/1\ s$ and $2.7 \times 10^{-17}/10^5\ s$. During the testing period, the phase drift remained less than 0.006 rad over a span of 5 days, as shown in Fig. 4(b). Due to temperature fluctuations in the laboratory, there is some non-ideal degradation in the Allan variance and phase drift. We can compare our results with the requirements of the frequency distribution system for SKA. For SKA, the frequency distribution system introduces a coherent loss of no more than 2% within an integration time of 100 s, and the phase drift does not exceed 1 rad within 10 minutes [36]. For the observation frequency of 20 GHz, the requirement for coherent loss can be converted into Allan deviation, i.e., $1 \times 10^{-12}/1\ s$ and $1.0 \times 10^{-14}/100\ s$ [37]. In comparison, the system is well-suited for the frequency distribution tasks of the radio array.

For vibration detection, we conducted tests at distances of ~25 km and ~45 km from the RPF. We apply sinusoidal vibrations of 10 Hz and 40 Hz at these two locations. The choice of using sinusoidal waves is due to their reproducibility and controllability. Experiments on detecting other forms of vibrations can be found in the previous work [35]. The phase signals detected at the RPF end are illustrated in Fig. 5(a) and 4(c). The phase fluctuation in the absence of vibration can be considered as the noise floor of the system. As shown in Fig. 5(a) and 5(c), the noise floor detected is less than $15 \times 2\pi \cdot rad$. The amplitude of minimum detectable vibrations corresponding to this noise floor is 3.75 μm. Furthermore, since the sampling rate of the detection signal is 5 MHz, this corresponds to a positioning resolution of 20 m.

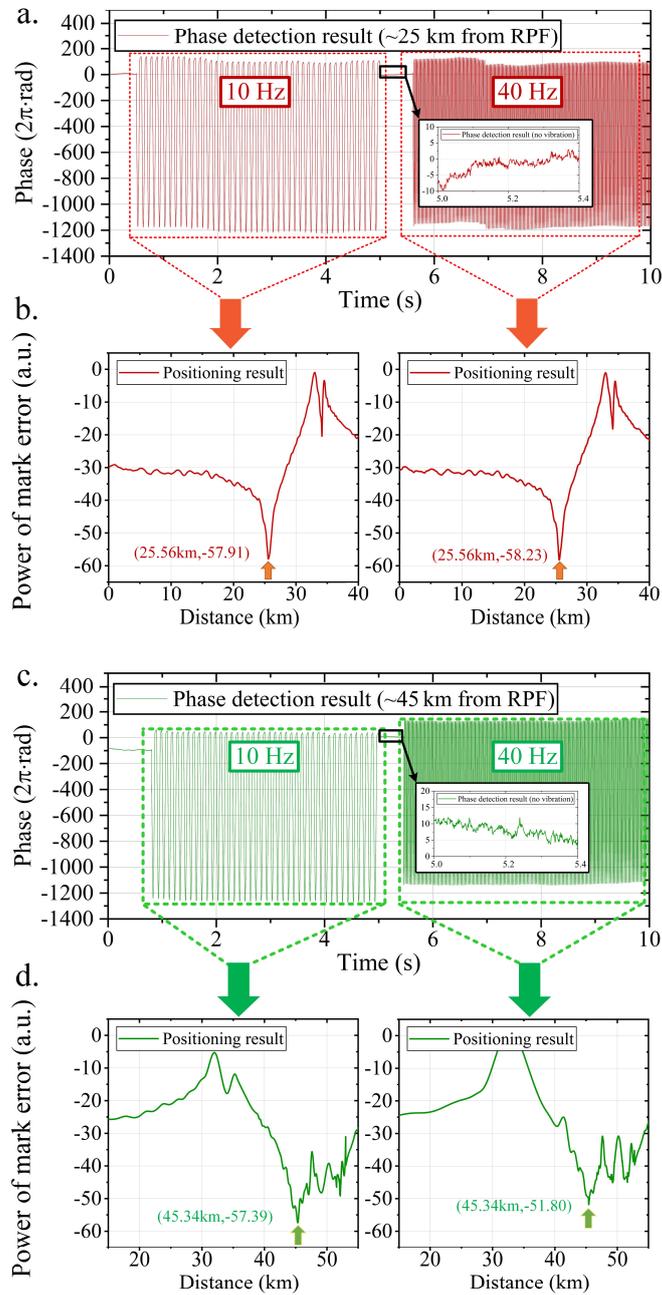

Fig. 5. Vibration detection test results. (a) Phase signal detected when vibration is applied 25 km away from the RPF. (b) The relationship curve between the power of mark error and distance for the vibration at the ~25-km position. (c) Phase signal detected when vibration is applied 45 km away from the RPF. (d) The relationship curve between the power of mark error and distance for the vibration at the ~45-km position.

Based on the phase data in Fig. 5(a) and 5(c), we calculate the specific location of the vibration. Localization results are shown in Fig. 5(b) and 5(d). After conducting multiple vibration experiments, the average positioning result for FS-1 is 25.556 km, with a positioning standard deviation of 22.93 m. For FS-2, the average positioning result is 45.344 km, with a positioning standard deviation of 24.14 m. We use the OTDR to verify the accuracy of the localization results. The OTDR measurement results for the FS placement locations are 25.558 km and 45.343 km, which match our actual localization results very well.

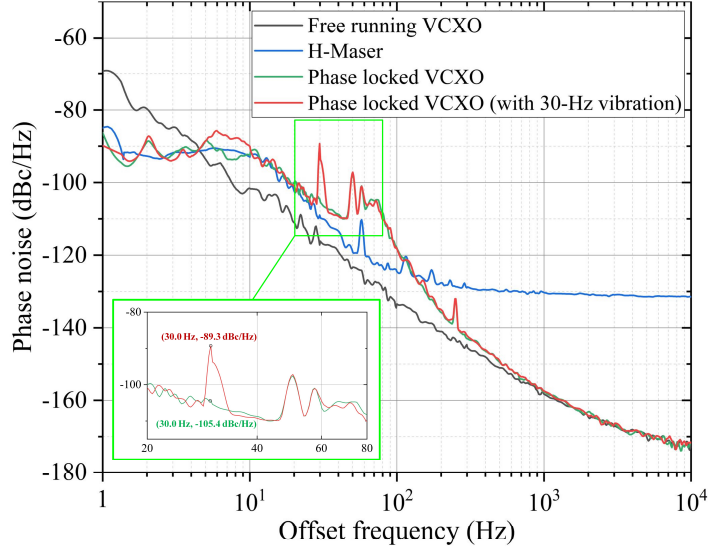

Fig. 6. Phase noise measurement results. The black line represents the phase noise curve of the free-running VCXO; The blue line shows the phase noise curve of the H-Maser; The green line represents the phase noise curve of the VCXO with phase locking under no vibration influence; The red line represents the phase noise curve of the phase-locked VCXO when a 30 Hz vibration is applied at 25 km.

As previously discussed, abnormal vibrations along the fiber link often indicate potential risk events. Although some risk events may not cause irreversible damage to the fiber network, they can impact the performance of frequency distribution. They primarily affect the phase noise of the synchronized frequency signal at the RPF. To visually demonstrate this impact, vibrations are applied at a distance of 25 km along the link. We use another phase noise analyzer, Rohde & Schwarz, FSWP26, to measure the phase noise of various 100-MHz signals: the free-running voltage-controlled crystal oscillator (VCXO) at the RPF end, the reference output from the H-maser, the phase locked VCXO at the RPF end, and the phase locked VCXO at the RPF end when 30-Hz vibrations are applied. Test results are shown in Fig.6. From the comparison of the black, blue, and green lines, the phase-locking effect of the frequency distribution system can be observed. However, when vibrations at frequencies of 30 Hz are applied, they affect the phase noise of the frequency signal at the RPF at the corresponding frequency offsets. The deterioration is due to the fact that the fiber-based frequency distribution system has an upper limit on its compensation capacity for fluctuations along the link [38]. Abnormal vibrations can exceed this upper limit, thereby affecting the phase noise. The peaks at 50-Hz and 250-Hz frequency offsets in Fig. 6 are caused by the 50-Hz common frequency of the grid and its harmonics. The peak near 60 Hz is due to the characteristics of the H-maser. The hybrid system we designed enables both frequency distribution and abnormal vibration detection. At the RPF, it will help users know which frequency offsets experience phase noise deterioration, allowing them to take proactive measures to mitigate the impact.

By analyzing the detection results, the time-domain characteristics of the vibrations and their locations can be determined. Previous studies have shown that it is feasible to identify the event type and assess the extent of the impact based on detection results [39]. Once these events are identified, the report about the frequency distribution system can be provided to RPF. If the event is likely to cause irreversible damage to the frequency distribution link, such as manual digging or mechanical excavation, maintenance personnel can be directed to perform risk assessment and maintenance based on the detected vibration location. If the event only negatively affects the performance of the frequency distribution system, RPF can determine whether the output by the system can still be used within the specified time based on the report.

## 4. Conclusion

This paper presents a hybrid fiber-based frequency-distribution and vibration detection system for large radio telescope arrays. The proposed system effectively addresses the dual challenges of maintaining precise frequency distribution and detecting and localizing vibrations within the optical fiber network. The system can monitor the reliability of the fiber network and mitigate potential risks posed by environmental disturbances without compromising frequency distribution performance. The experimental results indicate that the system meets the stringent frequency distribution requirements of large radio arrays, with good long-term stability and low phase drift. Furthermore, the capabilities of vibration detection and localization can provide valuable information for proactive maintenance and help ensure the reliability of fiber infrastructure. The success of this system provides valuable insights into the design and implementation of robust, hybrid systems for large-scale scientific observatories, such as SKA and ngVLA. Future work could explore further optimizations in vibration detection sensitivity and the integration of real-time monitoring capabilities to enhance overall performance.

**Funding.** National Natural Science Foundation of China (62171249); National Key Research and Development Program of China (2021YFA1402102); Tsinghua Initiative Scientific Research Program.

**Disclosures.** The authors declare no conflicts of interest.

**Data availability.** Data underlying the results presented in this paper are not publicly available at this time but may be obtained from the authors upon reasonable request.